\documentstyle[12pt]{article}
\newcommand{\PSbox}[3]{\mbox{\rule{0in}{#3}\includegraphics{#1}\hspace{#2}}}
\begin{document}

\def\T{{\cal T}}
\def\tr{{\rm Tr}}
\def \h {\hat}
\def \s {\sigma}
\def \p {\phi}
\def \ov {\over}
\def \fourth  {{1\ov 4} }
\def \four{{\textstyle {1\ov 4}}}
\def \a {\alpha}
\def\ep{\epsilon}
\def\vp {\varphi}
\def \ha{{\textstyle{1\over 2}}}
\def \td {\tilde }

\def\({\left( }
\def\){\right)}
\def\N{ {\cal N} }
\def\sei{ {\textstyle{1\ov 6} } }
\def\tre{ {\textstyle{1\ov 3} } }
\def\ads {$AdS_7$}
\def\adss{$AdS_7 \times S^4$}
\def\rr{4\pi g_s N}
\def\del{\partial }

\def\stop{.}
\def\comma{,}
\def\non{\nonumber}

%%%%%%%%%%%
 %macros here:
\newcommand{\norm}[1]{\raise.3ex\hbox{:} #1 \raise.3ex\hbox{:}\,}
\newcommand{\disc}{{\rm Disc}}
\def\gone{{}}
\def\T{{\cal T}}
\def\O{{\cal O}}
\def\tr{{\rm Tr}}
\def\det{{\rm det}}
\def\MeV{\ {{\rm MeV}}}
\def\GeV{\ {{\rm GeV}}}
\newcommand{\beq}{\begin{equation}}
\newcommand{\eeq}{\end{equation}}

\newcommand{\beqar}{\begin{eqnarray}}
\newcommand{\eeqar}{\end{eqnarray}}

\def\non{\nonumber}

\def\appendix{{\newpage\section*{Appendix}}\let\appendix\section%
        {\setcounter{section}{0}
        \gdef\thesection{\Alph{section}}}\section}

%With ssg.bst one needs this definition unless you are going to
%set up hyperlinking.
\def\href#1#2{#2}

%             SECTION SETUP

\catcode`\@=11
\def\theequation{\thesection.\arabic{equation}}
\@addtoreset{equation}{section}
\catcode`\@=12
\def\thetable{\thesection.\arabic{table}}
\def\thefigure{\thesection.\arabic{figure}}
%       References
%
\def\noj#1,#2,{{\bf #1} (19#2)\ }
\def\jou#1,#2,#3,{{\sl #1\/ }{\bf #2} (19#3)\ }
\def\ann#1,#2,{{\sl Ann.\ Physics\/ }{\bf #1} (19#2)\ }
\def\cmp#1,#2,{{\sl Comm.\ Math.\ Phys.\/ }{\bf #1} (19#2)\ }
\def\ma#1,#2,{{\sl Math.\ Ann.\/ }{\bf #1} (19#2)\ }
\def\jd#1,#2,{{\sl J.\ Diff.\ Geom.\/ }{\bf #1} (19#2)\ }
\def\invm#1,#2,{{\sl Invent.\ Math.\/ }{\bf #1} (19#2)\ }
\def\cq#1,#2,{{\sl Class.\ Quantum Grav.\/ }{\bf #1} (19#2)\ }
\def\cqg#1,#2,{{\sl Class.\ Quantum Grav.\/ }{\bf #1} (19#2)\ }
\def\ijmp#1,#2,{{\sl Int.\ J.\ Mod.\ Phys.\/ }{\bf A#1} (19#2)\ }
\def\jmphy#1,#2,{{\sl J.\ Geom.\ Phys.\/ }{\bf #1} (19#2)\ }
\def\jams#1,#2,{{\sl J.\ Amer.\ Math.\ Soc.\/ }{\bf #1} (19#2)\ }
\def\grg#1,#2,{{\sl Gen.\ Rel.\ Grav.\/ }{\bf #1} (19#2)\ }
\def\mpl#1,#2,{{\sl Mod.\ Phys.\ Lett.\/ }{\bf A#1} (19#2)\ }
\def\nc#1,#2,{{\sl Nuovo Cim.\/ }{\bf #1} (19#2)\ }
\def\np#1,#2,{{\sl Nucl.\ Phys.\/ }{\bf B#1} (19#2)\ }
\def\pl#1,#2,{{\sl Phys.\ Lett.\/ }{\bf #1B} (19#2)\ }
\def\pla#1,#2,{{\sl Phys.\ Lett.\/ }{\bf #1A} (19#2)\ }
\def\pr#1,#2,{{\sl Phys.\ Rev.\/ }{\bf #1} (19#2)\ }
\def\prd#1,#2,{{\sl Phys.\ Rev.\/ }{\bf D#1} (19#2)\ }
\def\prl#1,#2,{{\sl Phys.\ Rev.\ Lett.\/ }{\bf #1} (19#2)\ }
\def\prp#1,#2,{{\sl Phys.\ Rept.\/ }{\bf #1C} (19#2)\ }
\def\ptp#1,#2,{{\sl Prog.\ Theor.\ Phys.\/ }{\bf #1} (19#2)\ }
\def\ptpsup#1,#2,{{\sl Prog.\ Theor.\ Phys.\/ Suppl.\/ }{\bf #1}
(19#2)\ }
\def\rmp#1,#2,{{\sl Rev.\ Mod.\ Phys.\/ }{\bf #1} (19#2)\ }
\def\yadfiz#1,#2,#3[#4,#5]{{\sl Yad.\ Fiz.\/ }{\bf #1} (19#2) #3%
\ [{\sl Sov.\ J.\ Nucl.\ Phys.\/ }{\bf #4} (19#2) #5]}
\def\zh#1,#2,#3[#4,#5]{{\sl Zh..\ Exp.\ Theor.\ Fiz.\/ }{\bf #1}
(19#2) #3%
\ [{\sl Sov.\ Phys.\ JETP\/ }{\bf #4} (19#2) #5]}

%%%%%%%%%%%%%%%%%%%%%%%%%%%%%%%%%%%%%%%%%%%%% 
%%%%%%%%

\begin{titlepage}
\begin{center}
\vspace*{-3cm} \today \hfill UCB-PTH-98/52\\
hep-th/9810186 \hfill LBNL-42424\\
{}~{} \hfill CERN-TH/98-338\\
{}~{} \hfill Imperial/TP/98-99/5

\vskip 2cm

{\Large{\bf Large $N$ QCD from Rotating Branes}}

\vskip 1.5cm

{Csaba Cs\'aki$^{a,b,}$\footnote{Research fellow, Miller Institute for Basic
Research in Science.}, Yaron Oz$^c$,  Jorge  Russo$^{d,e}$ and John
Terning$^{a,b}$}

\vfil

  {\it ${}^a$ Department of Physics

     University of California, Berkeley, CA 94720, USA

\medskip

${}^b$ Theoretical Physics Group

     Ernest Orlando Lawrence Berkeley National Laboratory

     University of California, Berkeley, CA 94720, USA

\medskip

${}^c$ Theory Division, CERN

CH-1211, Geneva 23, Switzerland

\medskip

${}^d$ Theoretical Physics Group, Blackett Laboratory

  Imperial College,  London SW7 2BZ, U.K.

\medskip

    ${}^e$ SISSA, I-34013 Trieste, Italy}

\vfil

{\tt csaki@thwk5.lbl.gov, yaron.oz@cern.ch,  j.russo@ic.ac.uk,
terning@alvin.lbl.gov}

\vspace{5mm}

\begin{abstract}
We study large $N$ $SU(N)$ Yang-Mills theory in
three and four dimensions using a one-parameter
family of
supergravity models which originate from  non-extremal rotating
D-branes.
We show explicitly that varying this
``angular momentum"
parameter decouples the Ka\-lu\-za-Klein
modes associated with the compact D-brane coordinate,
while the mass ratios for ordinary glueballs are quite stable
against this variation, and are in good agreement with
the latest lattice results.
We also compute the topological susceptibility
and the gluon condensate
as a function of the ``angular momentum" parameter.

\end{abstract}
\end{center}
\end{titlepage}

\newpage

%%%%%%%%%%%%%%%%%%%%%%%%%%%%%%%%%
\section{Introduction}
%%%%%%%%%%%%%%%%%%%%%%%%%%%%%%%%%
\setcounter{figure}{0}
\setcounter{table}{0}
Generalizing the
conjectured duality \cite{mal} between
large $N$ superconformal field theories and superstring or M theory on
anti-de Sitter (AdS) backgrounds, Witten proposed
an approach to studying large $N$ non-supersymmetric theories such as pure QCD
using a dual supergravity (string theory)  description \cite{WittenAdsThermal}.
The basic idea is  to start with
 $d+1$ dimensional superconformal field theories at finite temperature
 -- thus breaking the superconformal invariance -- and obtain a
 $d$ dimensional  non-supersymmetric gauge theory
at zero temperature by dimensional reduction in the Euclidean time direction.
The AdS space is then replaced by a certain limit of the Schwarzschild
geometry describing a black hole in an AdS space.

When the curvature of the space is small compared to the string scale
(or, in the case of M theory, Planck scale),
supergravity provides an adequate effective
description that exhibits a qualitative agreement
with pure QCD in three and four dimensions
\cite{WittenAdsThermal,GO98}.
The supergravity limit of string theory
(i.e. infinite string tension, $\alpha'\to 0$ limit)  corresponds to
the strong coupling limit of the gauge theory ($\lambda=g_{YM}^2N \gg 1$), with
$1/\lambda $
playing the role of $\alpha '$.
In the approach of \cite{WittenAdsThermal}, the gauge theory has an
ultraviolet cutoff proportional to the temperature $T$;
the supergravity approximation should describe the large $N$ gauge theory
in the strong coupling regime with a finite  ultraviolet cutoff.
This is analogous to a strong coupling lattice gauge theory with lattice
spacing $\sim 1/T$ \cite{GO98}.
In the limit that the  ultraviolet cutoff is sent to
infinity,  one has to study the theory at small $\lambda$, and the
supergravity description breaks down.
To calculate the spectrum in this regime, a better understanding of
string theory with  Ramond-Ramond (R-R) 
background fields is required.

Glueball masses in the supergravity approximation have been computed in
\cite{COOT98,jevic}.
The Witten model \cite{WittenAdsThermal}
contains in  addition to the glueballs 
certain Kaluza-Klein (KK) particles
with masses of the order of the glueball masses. These KK modes
do not correspond to any states in the Yang-Mills theory, and therefore
they should decouple in the ``continuum" limit.
The KK states do not decouple
with the  inclusion of the leading $1/\lambda $ corrections
\cite{ORT}.
Although  such states can
decouple in a full string theory treatment,
there may be generalizations of the Witten model that
have a more direct connection with the continuum gauge theory already at
the supergravity level (at least in the sector of states with spin $\leq 2$
that can appear in a supergravity description).
A similar situation arises in  lattice gauge theory.
It is well known that the action one starts with
has a significant effect on the speed at which one gets to the continuum
limit.
One can add to the lattice action deformations which are irrelevant in
the continuum and arrive at an appropriate effective description of the
continuum theory  while having a larger lattice spacing (such a deformed 
action is called an ``improved" lattice action).
A similar strategy  in the dual supergravity picture would
correspond to a suitable modification of the background
metric, so as to have an
appropriate effective description of the gauge theory while still having
a finite  ultraviolet cutoff.
An important test of the proposal
is to check that the KK  modes in the supergravity
description that do not correspond to  gauge degrees of freedom
are heavy and decouple, and at the same time
the infrared physics is not significantly altered.
In this paper we make the first step in this direction by examining a
generalization of the Witten model that has an additional parameter.

A more general approach to the conjectured correspondence between gauge
theories and
M-theory  requires the investigation of  supergravity  compactifications
which asymptotically approach anti-de Sitter backgrounds, e.g.
 $AdS_5\times S^5$ or $AdS_7\times S^4$ (see e.g. \cite{GKP98,WittenAdS}).
There exist a few supergravity backgrounds that generalize the Witten model
and which are regular everywhere. These are essentially obtained by starting
with
the rotating version of  the non-extremal D4 brane
background (or rotating D3 brane background, in the case of ${\rm QCD}_3$)
and taking
a field theory limit as in \cite{mal}. These models were investigated in
\cite{russo}. The deformation of the background proposed in
\cite{WittenAdsThermal}
is parameterized by an ``angular momentum" parameter (the supergravity
background
is actually static, with the Euclidean time
playing the role of an internal angle).
In this paper we determine numerically the scalar 
and pseudoscalar spectrum of these models
as a function of the  angular momentum parameter and compare the results to 
those obtained by lattice calculations.
We also compute the gluon condensate and the topological susceptibility

The paper is organized as follows:
Section 2 is devoted to the study of supergravity models for pure
QCD in $3+1$ dimensions.
The models are described in  Section 2.1.
 In  Section 2.2 we  compute the scalar glueball mass spectrum
and analyze its dependence on the  angular momentum.
In  Section 2.3 we calculate  the mass spectrum of some  KK modes.
It will be shown that the KK modes associated with the compact
D-brane coordinate
decouple as  the angular momentum parameter
is increased. This, however, is not the case for
the $SO(3)$ non-singlet
KK modes with vanishing $U(1)$ charge in the compact D-brane coordinate.
In  Section 2.4  we compute the gluon condensate from the
free energy associated with the
supergravity background.
In  Section 2.5
 we compute the topological susceptibility
and its dependence on the  angular momentum parameter.
 Section 3 contains an analogous study for the case of  
QCD in $2+1$ dimensions. The
conclusions are similar in both cases, and they are summarized in Section 4.

%%%%%%%%%%%%%%%%%%%%%%%%%%%%%%%%%%%%%%%%%%%%
\section{QCD in $3+1$ dimensions}
%%%%%%%%%%%%%%%%%%%%%%%%%%%%%%%%%%%%%%%%%%%%
\setcounter{figure}{0}
\setcounter{table}{0}
%%%%%%%%%%
\subsection{Supergravity Models for ${\rm QCD}_4$}
%%%%%%%%%%%%%%

One way to construct non-supersymmetric models of QCD based on supergravity
is to start from the  non-extremal D4 brane metric, and
view the Euclidean time coordinate as an internal coordinate compactified on
a circle of radius $(2\pi T_H)^{-1}$ \cite{WittenAdsThermal}.
Possible generalizations of this proposal are constrained by
the no-hair theorem, which implies that the most general regular manifold
with only D4 brane charges (and an isometry group containing ${\bf R}^4$)
is given by the rotating version of the non-extremal D4 brane, which has
two additional  parameters representing angular momenta in two different
planes.
The Euclidean version of this metric  (related to the
 rotating M5 brane metric by dimensional reduction)
was used in \cite{russo} to construct models for QCD with extra parameters.
Here we  investigate in detail the case when there is  one non-vanishing
angular momentum, parametrized by $a$.
The field theory limit of the Euclidean rotating M5 brane with angular 
momentum
component in one plane is given by the metric \cite{russo}
\beqar
ds^2_{11}&=& \Delta^{1/3}\bigg[ {U^2\ov (\pi N)^{1/3} }
\bigg( \sum_{i=1}^5 dx_i^2 +\big( 1-{U_0^6\ov U^6\Delta }\big)  d\tau^2
\bigg)
\non\\
&+& (\pi N)^{2/3}\bigg(d\theta^2+{\td \Delta\ov \Delta} \sin^2\theta d\varphi^2
+{1\ov \Delta } \cos^2\theta d\Omega_2^2
\non\\
 &-&{2a^2 U_0^3\ov U^4\Delta (\pi N)^{1/2} } \sin^2\theta d\tau d\varphi
\bigg)
+ {4(\pi N)^{2/3}dU^2\ov U^2(1-{a^4\ov U^4}-{U_0^6\ov U^6})} \bigg]\ ,
\label{once}
\eeqar
where $x_1,\ldots,x_5$ are the coordinates along the M5 brane where the
gauge theory lives, $U$ is the ``radial" coordinate of the AdS space, while the
remaining four coordinates parameterize the angular variables of $S^4$, and
where
we have introduced
\beq
\Delta=1-{a^4\cos^2\theta \ov U^4}\ ,\ \ \ \ \td \Delta=1-{a^4\ov U^4} \ .
\eeq
Dimensional reduction along $x_5$ (which will play the role of the ``eleventh"
dimension) gives
$N$ rotating non-extremal D4 branes, which in the low energy regime should be
described by
$SU(N)$ Yang-Mills theory at finite temperature $T_H$, perturbed by some
operator
associated with the rotation.
The $3+1$ dimensional $SU(N)$ Yang-Mills theory at  zero-temperature can be
described
by making $x_4\to -i x_0$, and viewing
$\tau $ as parameterizing a space-like circle with radius $R_0=(2\pi
T_H)^{-1}$, where
fermions obey anti-periodic boundary conditions. At energies much lower
than $1/R_0$,
the theory is effectively $3+1$ dimensional.
Because of the boundary conditions, fermions
 and scalar particles get  masses proportional to the inverse radius,
so that, as $R_0\to 0$, they should decouple from the low-energy physics,
leaving  pure Yang-Mills theory as  low-energy theory.

The gauge coupling $g_4^2$ in the $3+1$ dimensional Yang-Mills theory is 
given
by the ratio between the periods of the eleven-dimensional coordinates $x_5$
and
$\tau $ times $2\pi $.
It is convenient to introduce ordinary angular coordinates
$\theta_1$, and $\theta_2$ which are $2\pi $-periodic by
\beq
\tau =R_0 \theta_2\ ,\ \ \ \ x_5={g_4^2\over 2\pi} R_0 \theta_1=
{\lambda \ov  N} R_0 \theta_1\ ,
\label{angg}
\eeq
\beq
R_0=(2\pi T_H)^{-1}={A\ov 3u_0}\  , \ \ \ \ \ \ \ A\equiv {u_0^4\ov u_H^4-\tre
a^4} \ ,
\label{auh}
\eeq
where $u_H$ is the location of the horizon, and
we have introduced the 't Hooft coupling
\beq
\lambda\equiv {g_4^2N\over 2\pi }\ ,
\eeq
% The gauge coupling $g_4^2$ in the $3+1$ dimensional Yang-Mills theory is given
%by the ratio between the periods of the eleven-dimensional coordinates $x_5$
%and
%$\tau $. It is convenient to introduce ordinary angular coordinates
%$\theta_1$ and $\theta_2$ which are $2\pi $-periodic by
%\beq
%\tau =R_0 \theta_2\ ,\ \ \ \ x_5=g_4^2 R_0 \theta_1=
%{\lambda \ov  N} R_0 \theta_1\ ,
%\label{angg}
%\eeq
%\beq
%R_0=(2\pi T_H)^{-1}={A\ov 3u_0}\  , \ \ \ \ \ \ \ A\equiv {u_0^4\ov u_H^4-\tre
%a^4} \ ,
%\label{auh}
%\eeq
%where $u_H$ is the location of the horizon, and
%we have introduced the 't Hooft coupling
%\beq
%\lambda\equiv g_4^2N\ ,
%\eeq
the coordinate $u$ by $U=2(\pi N)^{1/2} u$, and rescaled
$a\to 2(\pi N)^{1/2}a $.
By dimensional reduction in  $\theta_1$,
one obtains the metric
\beqar
ds^2_{\rm IIA}&=&{2\pi \lambda A \ov  3u_0} u \Delta ^{1/2}\bigg[ 4u^2
\big( -dx_0^2+dx_1^2+dx_2^2+dx^2_3\big)
\non\\
&+& { 4A^2\ov 9u_0^2} u^2 \ (1-{u_0^6\ov u^6 \Delta }) d\theta_2^2
+ {4\ du^2  \ov u^2 (1-{a^4\ov u^4}-{u_0^6\ov u^6 }) }
\non\\
&+& d\theta^2+{\td \Delta\ov \Delta} \sin^2\theta d\varphi^2
+{1\ov \Delta } \cos^2\theta d\Omega_2^2
 -{4a^2 A u_0^2\ov 3u^4\Delta } \sin^2\theta d\theta_2 d\varphi \bigg],
\label{pocho}
\eeqar
with a dilaton background
\beq
e^{2\phi }={8\pi\ov 27} {A^3\lambda^3 u^3\Delta^{1/2}\ov u_0^3} {1\ov N^2}
 \ \ .
\label{dilz}
\eeq
With this normalization, the metric reduces to Eq.~(4.8) of
ref.~\cite{WittenAdsThermal} after setting $a=0$.
The string coupling $e^\phi $ is of order $1/N$, and the metric has become
independent of $N$,
which is consistent with the expectation that in the large $N$ limit the string
spectrum
should be independent of $N$.  The metric is regular, and
 the location of the horizon is given by
\beq
u_H^6-a^4 u_H^2-u_0^6=0\ ,
\label{hhm}
\eeq
i.e.
\beq
u_H^2={a^4\ov \gamma u_0^2}+ \tre \gamma u_0^2\ ,\ \ \ \ \
\gamma=3\left(\ha+\ha \sqrt{1-{4\ov 27}\left({a\ov u_0}\right)^{12} }
\right)^{1/3}\ .
\label{uuhh}
\eeq
Note that for large $a$, one has the approximate expressions
\begin{eqnarray}
&& u_H^2\approx a^2 +{u_0^6\over 2 a^4}\ ,\ \ \ \ \ \ \ A\approx
\frac{3u_0^4}{2a^4} \ ,
\label{swws}
\end{eqnarray}
($u_H^2$ is always  real).
This shows, in particular,
that the radius $R_0 =A/(3 u_0)$ can be made very small by increasing  $a/u_0$.
This is essentially the mechanism that will make the corresponding
KK modes decouple at large $a/u_0$. At small $a$, such
KK states have masses of the same order as the masses of the lightest glueball
states.

The string tension is given by \cite{russo}
\beq
\sigma ={4\ov 3}\lambda A u_0^2=4\lambda {u_0^6\ov 3 u_H^4-a^4}\ .
\label{sstt}
\eeq
String excitations should have masses of order
$\sigma^{1/2}$.
The spin $\leq 2$ glueballs that remain in the supergravity approximation
-- whose masses are determined from the Laplace equation --
have masses which are independent of $\lambda $.
The supergravity approximation is valid for $\lambda A \gg 1$
so  that all curvature invariants are small \cite{russo}.
In this limit the  spin $> 2$ glueballs corresponding to
string excitations will be much heavier than the
supergravity glueballs.

%%%%%%%%%%%%%%%%%%%%%%%%%%%%%%%%%%%%%%%%%%%
\subsection{Spectrum of glueball masses}
%%%%%%%%%%%%%%%%%%%%%%%%%%%%%%%%%%%%%%%%%
The glueball masses are  obtained by computing correlation
functions of gauge invariant local operators or the Wilson loops, and
looking for particle poles. Following \cite{GKP98,WittenAdS},
correlation functions of local operators $\O$  are related at large $N$ and
large $g_{YM}^2N$ to tree level amplitudes of supergravity.
The generating functional for the correlation
function of $\O$ is the string partition function evaluated with
specified boundary values $\varphi_0$ of the string fields.  When the
supergravity description is applicable we have
\beq
\langle e^{-\int  d^4x \varphi_0(x) \O(x)} \rangle = e^{-I_{SG}(\varphi_0)}
\label{cor}
\comma
\eeq
where $I_{SG}$ is the supergravity action.

The spectrum of the scalar
glueball\footnote{In the following we will use the notation $J^{PC}$
for the glueballs, where $J$ is the glueball spin, and $P$, $C$ refer
to the parity and charge conjugation quantum numbers respectively.}
$0^{++}$ is obtained by finding the normalizable
 solution to the supergravity equation for the
dilaton mode $\Phi$ that couples to
 $\tr F^2$, which
is the lowest dimension operator
with $0^{++}$ quantum numbers.

The equation for $\Phi$ reads
\beq
\label{dil}
\del_\mu  [ \sqrt{g}e^{-2\phi }g^{\mu\nu}\del_\nu\Phi ]=0,
\eeq
where $g_{\mu\nu}$ is the string frame metric.

\begin{figure}
\PSbox{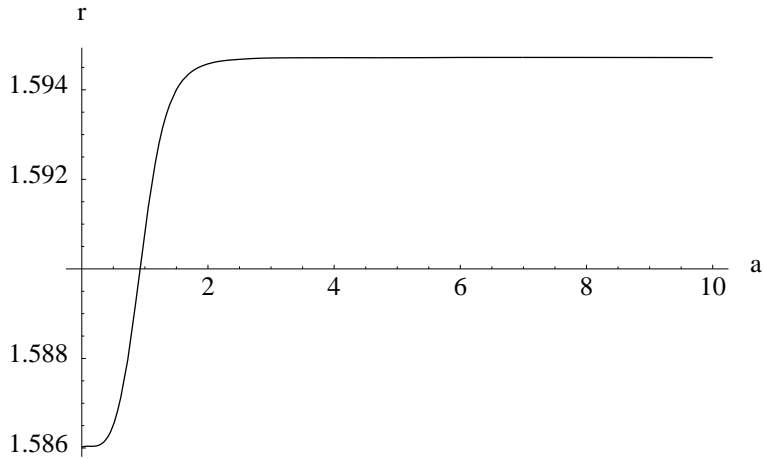 hscale=100 vscale=100 hoffset=50
voffset=-60}{13.7cm}{4.5cm}
\begin{center}
\parbox{5in}{\caption{The 
dependence of the ratio $r=\frac{M_{0^{++*}}}{M_{0^{++}}}$
of the masses of the
first excited ($0^{++*}$) glueball
state to the lowest $0^{++}$ glueball state on the parameter $a$ (in units
where $u_0=1$). The ratio
changes very little and takes on its asymptotic value quickly.
\label{fig:4ddila}}}
\end{center}
\end{figure}

We look for $\theta $-independent solutions of the form
$\Phi=\chi (u) e^{ik\cdot x}$. One obtains the equation
\beq
{1\ov u^3} \del_u [u (u^6-a^4 u^2-u_0^6) 
\chi' (u)]= -M^2 \chi (u)\ ,\ \ \ \ \ M^2=-k^2\ ,
\label{glus}
\eeq
where the eigenvalues $M$ are the glueball masses. The solution of this
(ordinary) differential equation has to be normalizable and regular
both at $u \to u_H$ and $u\to \infty$. The eigenvalues of this
equation can be easily obtained numerically \cite{COOT98}
by using the ``shooting" method.
One first finds the asymptotic behavior of $\chi (u)$ for large
$u$, and then numerically integrates this solution back to the horizon.
The solutions regular at the horizon will have a finite derivative
at $u_H$. This condition will determine the possible values of the
glueball masses $M$. The results of this numerical procedure
are presented in Table \ref{tab:4ddila} and in Figure \ref{fig:4ddila}.
One can see from Figure \ref{fig:4ddila}
that the ratios of the masses of the excited glueball states compared
to the ground state are very stable with respect to the variation
in the parameter $a$, even though both quantities themselves grow
like $M^2 \propto a^2$ for large $a$. The asymptotic value of the
mass ratios is taken on very quickly, $a/u_0 \approx 2$ is sufficiently
large to be in the asymptotic region.
A priori one could have expected that these mass
ratios may change significantly when $a$ is varied.
This leads one to suspect that there is a dynamical reason for the
stability of the ratios of masses.

\begin{table}[htbp]
\centering
\begin{tabular}{l|ccc}
state & lattice, $N=3$ &
supergravity  $a=0$ & supergravity $a\to \infty$ \\
 \hline
 $0^{++}$ & $1.61 \pm 0.15$   & 1.61 {\rm (input)} & 1.61 {\rm (input)} \\
 $0^{++*}$ &  $2.8 $  & 2.55 &  2.56 \\
 $0^{++**}$ &   - & 3.46  & 3.48 \\
 $0^{++***}$ &  -  & 4.36 &  4.40 \\
\end{tabular}
\parbox{5in}{\caption{Masses of the first few $0^{++}$ glueballs in
QCD$_4$, in GeV,
from supergravity compared
to the available lattice results. The first column gives the lattice result
\protect\cite{Teper97,MorningstarPeardon},
the second the supergravity result for $a=0$ while the third the
supergravity result in the $a\to \infty$ limit.
The authors of ref. \protect\cite{MorningstarPeardon} do note
quote an error on the preliminary lattice
result for $0^{++*}$. Note that the change
from $a=0$ to $a=\infty$ in the supergravity predictions is
tiny.\label{tab:4ddila}}}
\end{table}

Let us now consider the $ 0^{-+}$ glueballs. The lowest dimension
operator with $0^{-+}$ quantum numbers is
${\rm Tr}\, F\tilde F$.
On the D4 brane worldvolume,
the field that couples to this operator is the
R-R 1-form $A_\mu $.
In order to find the $ 0^{-+}$ glueball masses we have to solve its
equation of motion
\beq
\del_\nu\big[ \sqrt{g} g^{\mu\rho}g^{\nu\sigma } (\del_\rho A_\sigma
- \del_\sigma A_\rho ) \big]
=0
\ ,\ \ \ \ \mu ,\nu =1,...,10\ .
\label{mmxx}
\eeq
Consider solutions of the form
\beq
A_{\rm \theta_2}=\chi_{\theta _2}(u)\ e^{ik\cdot x}\ ,\ \ \ \ \ \ \
A_\mu=0\ \ {\rm if}\ \
\mu\neq\theta_2\ .
\label{RReq}
\eeq
Plugging this into (\ref{RReq}),
we obtain
\beq
\del_\nu \big[\sqrt{g} g^{\theta_2\theta_2} g^{\nu\sigma }\del_\sigma
A_{\theta_2}
\big]=0 \ ,
\eeq
which reads
\beq
{1\ov u^5} (u^6-a^4u^2-u_0^6)\del_u
\big[u^3(u^4-a^4)\chi_{\theta_2}'(u)\big]
=-M^2 (u^4-a^4) \chi_{\theta_2}(u)
\stop
\label{0-+}
\eeq
For $a=0$, it yields  Eq. (2.9) of ~\cite{haoz}, as required.
When $a\neq 0$ there are no solutions of the form (\ref{RReq}).
The reason is that  the $g_{\theta_2\varphi}$
component of the metric
(\ref{pocho}) is non-vanishing for $a\neq 0$ and, as a result,
the $\varphi $ component of the Maxwell equation is not
satisfied automatically (solutions contain a non-vanishing
component $A_\varphi $).

We will work in the approximation $a/u_0\gg 1$.
In this approximation the non-diagonal $g_{\theta_2\varphi}$
part of the metric can be neglected, and there are solutions of the form
(\ref{RReq}).
Effectively, we can solve (\ref{0-+}) in the limit $a\gg u_0$.
We must however keep in mind that we need $u_0\neq 0$ to regularize the
horizon, and the actual limit that is taken is
$a/u_0$ large at fixed $u_0$ (so that curvature invariants are bounded from
above and they are small for sufficiently large
't Hooft coupling $\lambda $).

\begin{figure}
\PSbox{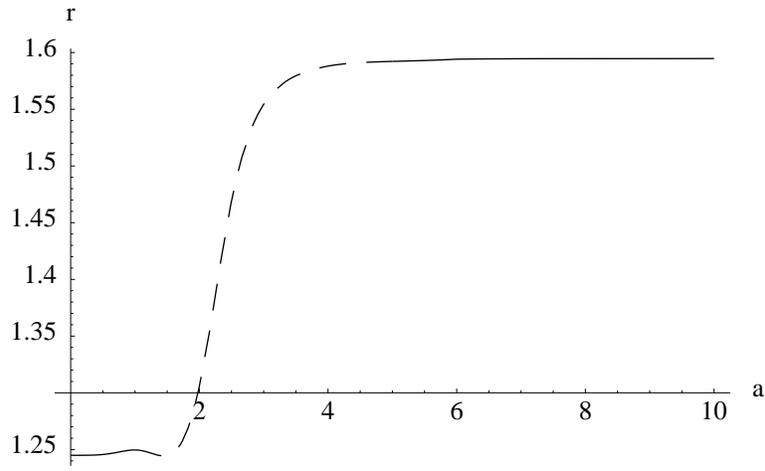 hscale=100 vscale=100 hoffset=50
voffset=-60}{13.7cm}{4.5cm}
\begin{center}
\parbox{5in}{\caption{The dependence of the ratio 
$r=\frac{M_{0^{-+}}}{M_{0^{++}}}$
on the parameter $a$ (in units
where $u_0=1$). The
change in the ratio is stable against the variation of $a$, however it
increases by about 25\% while going to $a=\infty$.
The change is in agreement with lattice simulations. As explained in the 
text, this figure is reliable only for the 
regions $a \ll u_0$ or $a \gg u_0$ which are shown in the plot with a solid 
line, while for the intermediate region denoted by a dashed line there are 
corrections due to the non-vanishing off-diagonal component of the
metric $g_{\theta_2 \varphi}$.
\label{fig:0-+}}}
\end{center}
\end{figure}

The mass spectrum from  (\ref{0-+}) can be obtained using a similar
numerical method as for the $0^{++}$ glueballs. The dependence of
the lightest $0^{-+}$ glueball mass on $a$ is presented in Figure
\ref{fig:0-+},
whereas the $0^{-+}$ glueball mass spectrum in Table \ref{tab:0-+}.
Note that while  masses ratios are fairly stable against the variation
of $a$ (they again grow like $M^2 \propto a^2$), the actual values
of the mass ratios compared to $0^{++}$ increase by a sizeable ($\sim 25\%$)
value. The change is in the right direction as suggested by recent
improved  lattice simulations \cite{Peardon}.  The mass of the second
$0^{-+}$ state also increases and is in agreement with the new lattice
results \cite{Peardon}.

\begin{table}[htbp]
\centering
\begin{tabular}{l|ccc}
state & lattice, $N=3$ &
supergravity  $a=0$ & supergravity $a\to \infty$ \\
 \hline
 $0^{-+}$ & 2.59 $\pm$0.13   & 2.00 & 2.56 \\
 $0^{-+*}$ &  3.64 $\pm$0.18   & 2.98 &  3.49 \\
 $0^{-+**}$ &   - & 3.91  & 4.40 \\
 $0^{-+***}$ &  -  & 4.83 &  5.30 \\
\end{tabular}
\parbox{5in}{\caption{Masses of the first few $0^{-+}$ glueballs in
QCD$_4$, in GeV,
from supergravity compared
to the available lattice results. The first column gives the lattice result,
the second the supergravity result for $a=0$ while the third the
supergravity result in the $a\to \infty$ limit. Note that the change
from $a=0$ to $a=\infty$ in the supergravity predictions is sizeable,
of the order $\sim 25 \%$.\label{tab:0-+}}}
\end{table}

We can directly compare the ratio of masses of the lowest glueball states 
$0^{-+}$ and $0^{++}$ with lattice results 
\cite{Teper97,MorningstarPeardon,Peardon}.  Since one of the  largest 
errors in the lattice calculation of glueball masses comes from setting 
the overall scale\footnote{We thank M. Peardon for emphasizing this point 
to us.} the ratios of masses are even more
accurately known from the lattice than the masses themselves.  
Using the lattice results \cite{MorningstarPeardon,Peardon,Michael}
in the more accurate ``lattice units'' $r_0$:
\beq r_0 M_{0^{++}}=4.33\pm 0.05, 
\ \ \ r_0 M_{0^{-+}}=6.33\pm 0.07, \ \ \ r_0 M_{0^{-+*}}=8.9\pm 0.1,\eeq
we find:
\beqar
\label{que}
&\left(\frac{M_{0^{-+}}}{M_{0^{++}}}
\right)^{a=0}_{{\rm supergravity}} &= 1.24  
\non\\
&\left(\frac{M_{0^{-+}}}{M_{0^{++}}}
\right)^{a\rightarrow \infty}_{{\rm supergravity}} &= 1.59  
\non\\
&\left(\frac{M_{0^{-+}}}{M_{0^{++}}}\right)_{{\rm lattice~~~~~~}} &=
1.46\pm 0.03 
\eeqar

\beqar
\label{que2}
&\left(\frac{M_{0^{-+*}}}{M_{0^{++}}}\right)^{a=0}_{{\rm supergravity}}&=  
1.85
\non \\
&\left(\frac{M_{0^{-+*}}}{M_{0^{++}}}\right)^{a
\rightarrow \infty}_{{\rm supergravity}}&= 2.17
\non \\
&\left(\frac{M_{0^{-+*}}}{M_{0^{++}}}\right)_{{\rm lattice~~~~~~}} &= 
2.06\pm 0.05
\eeqar
One can see that taking the $a \rightarrow \infty$ improves the agreement
between the supergravity and lattice predictions significantly.  
One should however keep in mind that the supergravity results presented 
here are for the limit $N\to \infty$ and $\lambda \to \infty$, while
the lattice results are for $N=3$ and $\lambda$ small.\footnote{For example,
in Ref.~\protect\cite{MorningstarPeardon} the results are 
extrapolated to $\lambda =0$ from calculations in the region
$ 7.5 < g^2N <10$.} Direct lattice calculations for the large $N$ limit are
have just started to become available~\cite{Dalley}, however no reliable direct
estimate for the mass of the $0^{-+}$ is known yet.

%%%%%%%%%%%%%%%%%%%%%%%%%%%%%%%%%%%%%%%%%%%%

%%%%%%%%%%%%%%%%%%%%%%%%%%%%%%%%%%%%%%%%%%%
\subsection{Masses of Kaluza-Klein states}
%%%%%%%%%%%%%%%%%%%%%%%%%%%%%%%%%%%%%%%%%

\begin{figure}
\PSbox{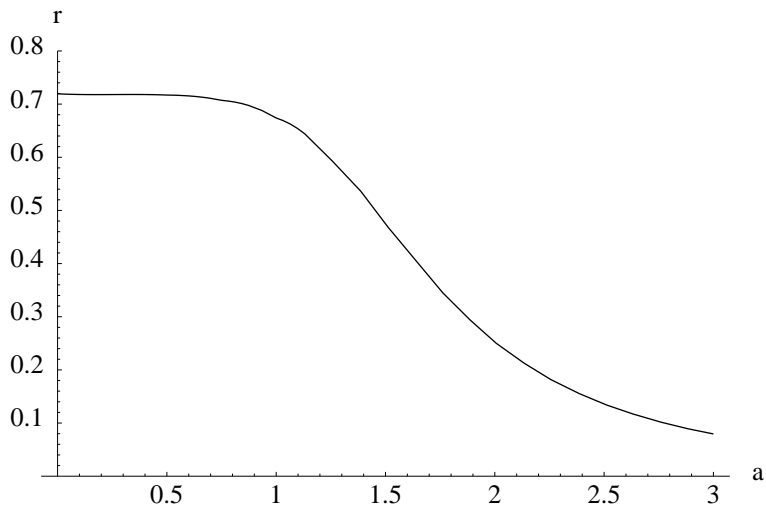 hscale=100 vscale=100 hoffset=50
voffset=-60}{13.7cm}{4.5cm}
\begin{center}
\parbox{5in}{\caption{The dependence of the ratio $r=\frac{M_{0^{++}}}{M_{KK}}$
of the  the lowest $0^{++}$ glueball
state compared to the KK mode wrapping the
$\theta_2$ circle on the parameter $a$ in units where $u_0=1$.
This KK mode decouples very quickly  from the spectrum even in
the supergravity approximation.  
\label{fig:KKdec}}}
\end{center}
\end{figure}

In the
supergravity approximation the $a=0$ model contains  additional light
KK modes in the spectrum whose masses are of the same order as
those of the glueball states \cite{ORT}.
In this section we investigate whether the additional parameter
of the model considered here can be tuned to decouple the KK modes already at
the
supergravity tree-level. In the following, it will be shown that this is indeed
the
case for
 the KK modes wrapped around the $\theta_2$ direction, which become very heavy
for $a/u_0 \gg 1$.
We thus look  for solutions of the dilaton equation (\ref{glus}) of the
form
\beq
\Phi=\chi (u)\  e^{ik\cdot x} e^{in\theta_2}\ .
\label{hhi}
\eeq
One finds the following equation:
\beq
{1\ov u^3} \del_u\bigg( u (u^6-a^4u^2-u_0^6)\chi'(u)\bigg)=
\bigg(-M^2+ {n^2\over R_0^2} {u^4-a^4\ov u^4-a^4-{u_0^6\ov u^2}} \bigg)
\ \chi(u)\ \ ,
\label{hhj}
\eeq
where $R_0$ is given in Eq.~(\ref{auh}).
This generalizes (\ref{glus}) to the case $n\neq 0$.
 We want to compare $M_0\equiv M(n=0)$ with $M_{\rm KK}\equiv M(n=1)$.
The question is how $M_0/M_{\rm KK}$ behaves as
a function of $a$ (we can set $u_0=1$).
The extra term proportional to $1/R_0^2$ gives a positive contribution to
the mass, so that $M_{\rm KK}$ should increase as $M^2_{KK}\propto a^8$
as $a$ is increased
(the KK radius $R_0=A/3u_0$
shrinks to zero as $a\to\infty $).
Thus one expects that $M_0/M_{\rm KK}\propto 1/a^3\to 0$
as $a$ increases. The numerical values of the masses of these KK modes
are displayed in Table \ref{tab:KKdec} and  Figure \ref{fig:KKdec}.
Note that the numerical evaluation of the masses of these KK modes
becomes more and more difficult as $a$ increases. This is because
the term with $1/R_0^2$ causes an overall shift of the masses, while the
splittings between the excited KK modes still remain of the
same order as for the ordinary glueballs.
As a result,  the solutions become
more and more quickly oscillating as $a$ increases, making numerical
treatments increasingly difficult. For this reason we display only
values up to $a/u_0=3$.

\begin{table}[htbp]
\centering
\begin{tabular}{l|cc}
state &
value for $a=0$ & value for $a = 3$ \\
 \hline
 $KK$ & 2.24 & 20.25 \\
 $KK^{*}$ & 3.12 &  20.37 \\
 $KK^{**}$ & 4.01 & 20.52\\
 $KK^{***}$ & 4.89 &  20.72\\
\end{tabular}
\parbox{5in}{\caption{Masses of the KK
modes which wrap around the $\theta_2$ circle and
have no corresponding states in
QCD$_4$, in GeV. The first column gives the masses
for $a=0$ while the second the
masses for $a=3$. Note that even for $a=3$ these states are heavier by
a factor of 10 than the $0^{++}$ glueball mass and are effectively
decoupled from the spectrum even in the supergravity limit.\label{tab:KKdec}}}
\end{table}

Above we have demonstrated that the KK modes which correspond to states
that wrap the $\theta_2$ direction  are effectively
decoupled from the spectrum even in the supergravity approximation.
However, there are other KK  modes in this theory,
and one would like to know whether these are decoupled as well.
The reason for the decoupling of the modes on $\theta_2$ is clear:
the radius of this direction shrinks to zero when
$a\to \infty$. However, the radii of the other compact directions
do not behave similarly. Therefore it is reasonable to expect that these
states will not decouple at the level of supergravity from
the spectrum (but they could decouple once  string theory
corrections are incorporated). We now demonstrate
by explicit calculation of the corresponding mass spectrum
 that this is indeed the case.

Consider  non-singlet modes which are independent of 
$\theta_2$, of the form
 $f(u) e^{ik\cdot x} \cos\varphi\sin\theta $. This corresponds to
spherical harmonics on $S^4$ with angular momentum $l=1$.
Plugging this ansatz into the dilaton equation (\ref{dil}) we find that
$f(u)$  satisfies the
equation ($u_0=1$)
\beq
u^3 \partial_u [(u^7-a^4u^3-u)f'(u)] -f(u) \left( k^2+
\frac{4u^2(4+3a^4u^2-4u^6)}{1+a^4u^2-u^6} \right)=0\ .
\eeq

\begin{figure}
\PSbox{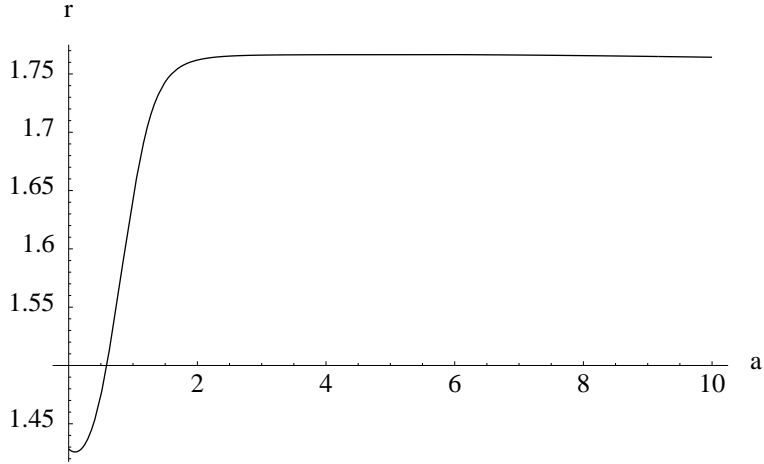 hscale=100 vscale=100 hoffset=50
voffset=-60}{13.7cm}{4.5cm}
\begin{center}
\parbox{5in}{\caption{The dependence of the ratio $r=\frac{M_{KK}}{M_{0^{++}}}$
of KK modes (corresponding to spherical harmonics with $l=1$ on $S^4$)
compared to the lowest $0^{++}$ glueball
state on the parameter $a$ in units where $u_0=1$.
This KK mode does not decouple from the spectrum in
the supergravity approximation even in the $a\to \infty$ limit.
\label{fig:KKnon}}}
\end{center}
\end{figure}

The results of the numerical analysis of the eigenvalues are presented in
Table \ref{tab:KKnon} and Figure \ref{fig:KKnon}. One can see that
these states do not decouple from the spectrum at the
supergravity level, instead their masses remain comparable to the
ordinary glueball masses.

\begin{table}[htbp]
\centering
\begin{tabular}{l|cc}
state &
value for $a=0$ & value for $a\to \infty$ \\
 \hline
 $KK$ & 2.30 & 2.84 \\
 $KK^{*}$ & 3.29 &  3.80 \\
 $KK^{**}$ & 4.23 & 4.74 \\
 $KK^{***}$ & 5.15 &  5.65 \\
\end{tabular}
\parbox{5in}{\caption{Masses of the KK
modes corresponding to $l=1$ angular momentum on the
$S^4$, in GeV. The first column gives the masses
for $a=0$ while the second the
masses in the $a\to \infty$ limit. Note that the change
from $a=0$ to $a=\infty$ in the supergravity predictions is not
sufficiently large in order to decouple these particular
states from the spectrum in the supergravity limit.\label{tab:KKnon}}}
\end{table}

%%%%%%%%%%%%%%%%%%%%%%%%%%%%%%%%%%%%%%%%%%%
\subsection{Free energy and Gluon condensation}
%%%%%%%%%%%%%%%%%%%%%%%%%%%%%%%%%%%%%%%%%

The standard relation between the thermal
partition function and free energy  $Z(T)=\exp(-F/T)$ relates the
free energy associated with the supergravity background
to the expectation value of the operator $\tr F_{\mu\nu}^2 $.
This relation was exploited in \cite{haoz}
to obtain a  prediction for the gluon condensate in the Witten ($a=0$)
supergravity model.
Let us now derive the corresponding supergravity result for general $a$.
{}From the rotating M5 brane metric (given in Eq.~(3.1) of \cite{russo}),
one can obtain the following
formulas for the ADM mass, entropy and angular momentum
(see also \cite{horsen,cvet}):
\beqar
M_{\rm ADM}&=&{V_5V(\Omega_4)\ov 4\pi G_N} 2m (1+{3\ov 4}\sinh^2\alpha )\ ,\
\  \ \ \ V(\Omega_4)={8\pi^2\over 3}\ ,
\label{madm}\\
S &=& {V_5V(\Omega_4)\ov 4 G_N} 2m r_H \cosh\alpha \ ,
\label{entro}\\
 J_H &=&{V_5 V(\Omega_4)\over 4\pi G_N} m l\cosh\alpha \ ,
\label{jjgg}
\eeqar
\beq
G_N={\kappa_{11}^2\ov 8\pi}=2^4\pi^7l_P^9\ ,
\label{ggn}
\eeq
where $G_N$ is Newton's constant in 11 dimensions, and $l_P$ is
the 11 dimensional Planck length.
The (magnetic) charge $N$ is related to $\alpha $ and $m$ by
\beq
2m \cosh\alpha\sinh \alpha=\pi Nl_P^3\ .
\eeq
The Hawking temperature and angular velocity are given by
\beq
T_H={3r_H^2+l^2\over 8\pi m \cosh\alpha }\ ,\ \ \ \ \
\Omega_H={l r_H\over 2m\cosh\alpha }\ .
\label{ppo}
\eeq
These quantities satisfy the first law of black hole thermodynamics:
\beq
dM_{\rm ADM}= T_H dS+ \Omega_H dJ_H\ .
\label{fll}
\eeq
We are interested in the field theory limit $l_P\to 0$, obtained by rescaling
variables as follows
\beq
r=u^2 l_P^3  (4\pi N)
\ ,\ \ \ m={1\over 2} u_0^6 l_P^9 (4\pi N)^3\ ,\ \ \ \ l=ia^2  l_P^3 (4\pi N)
\ .
\eeq
We get
\beq
E\equiv M_{\rm ADM}-M_{\rm extremal}={5\over 3\pi ^3} V_5 N^3  u_0^6 \ ,
\label{quno}
\eeq
\beq
T_H={3 u_0\ov  2\pi A}\
\ ,\ \ \ \ \ S={4\over 3\pi ^2} V_5 N^3 u_H^2 u_0^3 \ ,
\label{qdos}
\eeq
\beq
\Omega_H=i {2a^2 u_H^2\over u_0^3}
\ ,\ \ \ \ \  J_H=i{2\over 3\pi ^3} V_5 N^3  a^2 u_0^3\ ,
\label{qtres}
\eeq
with
\beq
A={3 u_0^4\over 3u_H^4-a^4}\ ,\ \ \ \ u_H^4-a^4={u_0^6\over u_H^2}\ .
\eeq
The free energy is given by
\beq\
F=E-T_HS- \Omega_H J_H=- {V_5\over 3\pi ^3}  N^3 u_0^6 \ .
\label{libre}
\eeq
For $a=0$ this reproduces the result of \cite{kletse} ($u_0= 2\pi A T_H/3$).
The M5 brane coordinate $x_5$ is
compactified on a circle with radius $R_0\lambda/N$, given by Eq. (\ref{auh}),
so that
\beq
V_5={V_4 \lambda\over T_H N}\ .
\eeq
The gluon condensate is then given by
\beq
\langle {1\over 4g_{\rm YM}^2}\tr\  F^2_{\mu\nu}(0) \rangle =
 - {F\over V_4 T_H}={4\over 27\pi }\lambda N^2 u_0 ^4 A^2\ .
\label{ssw}
\eeq
For $a=0$ this reduces to the corresponding result in \cite{haoz}
(setting $A=1$ and $u_0=2\pi T_H/3 $).
Expressing $u_0 $ in terms of the  string tension (\ref{sstt})
we obtain
\beq
\langle {1\over 4g_{\rm YM}^2} \tr\ F^2_{\mu\nu}(0) \rangle = {1\over 12\pi }
{N^2\over \lambda }
\sigma^2
\label{zzzz}
\stop
\eeq
Note that this relation is independent of $a$ (in particular, it applies to the
 $a=0$ case as well).
It has the expected dependence on $N$, and a
simple dependence on $\lambda $.

%%%%%%%%%%%%%%%%%%%%%%%%%%%%%%%%%%%%%%%%%%%
\subsection{Topological Susceptibility}
%%%%%%%%%%%%%%%%%%%%%%%%%%%%%%%%%%%%%%%%%

The topological susceptibility $\chi_t$ is defined by
\begin{equation}
\chi_t=\frac{1}{(16\pi^2)^2} \int d^4x \langle {\rm Tr}\, F\tilde{F}(x)
  {\rm Tr}\, F\tilde{F}(0)\rangle\ .
  \label{topol}
\end{equation}
The topological susceptibility measures the fluctuations of the
topological charge of the vacuum.  At large $N$ the Witten-Veneziano
formula \cite{Witteneta, Veneziano} relates the mass $m_{\eta'}$ in $SU(N)$
with $N_f$ quarks to
the topological susceptibility of $SU(N)$ without quarks:
\beq m_{\eta'}^2 = \frac{4N_f}{f_{\pi}^2} \chi_t \ .
\label{mass}
\eeq

The effective low-energy four dimensional brane theory
contains the coupling
\beq
 {1 \over 16 \pi^2 } \int d^4 x d\theta_2\,  A_{\theta_2}\
\tr \ F_{\mu \nu} F_{\lambda \sigma}
\epsilon^{\mu \nu \lambda \sigma} \ \comma
\label{I2}
\eeq
where $A_{\theta_2}$ is the component along the coordinate $\theta_2$
of the R-R 1-form $A_{\mu}$. We will consider zero mode ($M^2=0$)
configurations where $A_{\theta_2}$ is independent of
the world-volume coordinates.
Comparing to the standard Yang-Mills coupling,
\beq
{1\over 16 \pi^2}  \int d^4x\, \hat \theta
\ \tr \ F \tilde{F}\ \comma
\label{YM}
\eeq
one obtains  the relation
\beq
{\hat \theta } = \int_0^{2\pi }d\theta_2 \ A_{\theta_2}=2\pi A_{\theta_2}\
\stop
\label{dep}
\eeq

The action of the R-R 1-form is given by
\begin{equation}
I=\frac{1}{2\kappa_{10}^2} \int d^{10}x \sqrt{g} \frac{1}{4}
(\partial_{\mu} A_{\nu} - \partial_{\nu} A_{\mu})(\partial_{\mu'} A_{\nu'}
- \partial_{\nu'} A_{\mu'})g^{\mu\mu'}g^{\nu\nu'}\ .
\end{equation}
As discussed in Sect. 2.2, in the approximation that $a/u_0$ is either very
large or  very small,
the metric is diagonal and there are zero mode solutions of the form
 $A_{\theta_2}=A_{\theta_2}(u)$, $A_\mu=0\ ,\ \mu\neq\theta_2$.
The action reduces to
\begin{equation}
I=\frac{1}{4\kappa_{10}^2} \int d^{10}x \sqrt{g} \
(\frac{dA_{\theta_2} (u)}{du})^2 g^{uu}g^{\theta_2\theta_2}\ .
\end{equation}
Using Eq.~(\ref{pocho}) and integrating over the angular coordinates, this
becomes
\begin{equation}
I={2^7\pi^6A^2\lambda^3\over 27u_0^2\kappa_{10}^2} V_4
\int_{u_H}^\infty du\ u^3(u^4-a^4)(\frac{dA_{\theta_2} (u)}{du})^2\ .
\label{acci}
\eeq
The equation of motion is then given by
\begin{equation}
\partial_u [u^3(u^4-a^4)\partial_u A_{\theta_2}]=0\ .
\end{equation}
Whence
\beq
\del_u A_{\theta_2}= 6A^\infty_{\theta_2} C(a) {u_0^6\over u^3(u^4-a^4)}\ ,\ \
\
\eeq
\beq
\ \ A_{\theta_2}= A^\infty_{\theta_2}\bigg[1+ 6 C(a) \bigg(
{u_0^6\over 2 a^4u^2}+{u_0^6\over 4 a^6}\log {u^2-a^2\ov u^2+a^2}\bigg)\bigg]\
\label{sol1}
\stop
\eeq
The integration constant $C(a)$ will be fixed by assuming that
$A_{\theta_2}(u)$
vanishes at the horizon \cite{Wittentheta}. This gives
\beq
{1\over C(a)}= - {3u_0^6\over a^4u^2_H}+{3u_0^6\over 2 a^6}\log
{u^2_H+a^2\ov u^2_H-a^2}\ .
\label{cca}
\eeq
The other integration constant $A^\infty_{\theta_2}$ is related to the
$\hat\theta $-parameter by (\ref{dep}),  $2\pi A^\infty_{\theta_2}=\hat\theta
$.
Note that in the limit $a= 0$ one gets $C(0)=1$ and
$A_{\theta_2}(u)=A_{\theta_2}^\infty\big(1-{u_0^6\over u^6}\big)$.

Using $\kappa_{10}^2=\kappa_{11}^2/2\pi =2^6\pi^7 $ and $u_0=2\pi AT_H/3$, we
obtain
\beq
I=\hat\theta^2\ V_4{16 \pi \over 729} A^6 C(a)  \lambda^3  T^4_H \ .
\label{vvee}
\eeq
The topological susceptibility (\ref{topol})
can then be obtained by differentiating twice with respect to
$\hat \theta $:
\beq
\chi_t = {32 \pi \over 729} A^6 C(a)  \lambda^3  T^4_H \ ,
\label{otra}
\eeq
or, in terms of the string tension (\ref{sstt}),
\beq
\chi_t = {C(a) \over 8\pi ^3} \lambda \sigma^2\ .
\label{kkh}
\eeq
The $\hat\theta$-dependence of the vacuum energy of the form $\hat\theta^2$
is the result anticipated in \cite{Wittentheta} for the
$a=0$ model, and Eq.~(\ref{vvee}) shows that it holds 
for large $a$ too. In the large $N$ limit, this must be the case for
consistency
\cite{Wittentheta}.
For $a=0$, one has $A=1=C(a)$,  and Eq.~(\ref{otra}) reproduces the result
obtained in \cite{haoz}. In the large $a$ limit we have (see Eq.~(\ref{swws}))
\beq
C(a)\approx {a^6\ov 9 u_0^6\log {a\ov u_0}}\ ,
\non
\eeq
so that
\beq
\chi_t \approx {1\ov 18\pi^3} \lambda^3 {u_0^6\ov a^2 \log {a\ov u_0} }
={1\over 72\pi^3} {a^6\ov u_0^6 \log {a\ov u_0} }\lambda \sigma^2.
\label{ssl}
\eeq
This decreases if we increase $a/u_0$ at fixed $\lambda $ and $u_0$.

%%%%%%%%%%%%%%%%%%%%%%%%%%%%%%%%%%%%%%%%%
\section{QCD in $2+1$ dimensions}
%%%%%%%%%%%%%%%%%%%%%%%%%%%%%%%%%%%%%%%%%
\setcounter{figure}{0}
\setcounter{table}{0}
%%%%%%%%%%
\subsection{Supergravity Models for $QCD_3$}
%%%%%%%%%%%%%%

Analogous models for ${\rm QCD}_3$ can be obtained by starting with the
Euclidean rotating D3 brane (Eq.~(3.16) in \cite{russo} with
$x_0\to -i\tau$, $l\to il$)
and taking $\alpha' \to 0$ by rescaling variables as follows:
\beq
r=U\alpha '\ ,\ \ \ \ \ 2m=U_0^4 {\alpha '}^4\ ,\ \ \ \ l=a\alpha'\ .
\label{resc}
\eeq
In  the limit $\a'\to 0$ at fixed $U,a, U_0$ we obtain
\beqar \label{3dmetric}
&&ds^2_{\rm IIB} = \a' \Delta_0^{1/2} \bigg[{U^2\ov \sqrt{\rr }}\big(
h_0 d\tau^2 +dx_1^2+dx_2^2+dx_3^2\big)+ {\sqrt{\rr}\ dU^2\ov U^2(1 -{a^2\ov
U^2}
-{U_0^4\ov U^4})}
\nonumber \\
&&+\ \sqrt{\rr }\bigg( d\theta^2+ {\tilde \Delta_0\ov\Delta_0} \sin^2\theta
d\varphi^2
+{\cos^2\theta \ov\Delta_0} d\Omega_3^2 \bigg)
\ - \ {2 a U_0^2\ov U^2\Delta_0} \sin^2\theta d\tau d\varphi\bigg] \ ,
\nonumber \\
\eeqar
where
\beq
h_0=1-{U_0^4\ov U^4\Delta_0}\ ,\ \ \ \Delta_0=1 - {a^2 \cos^2\theta\ov U^2}\ ,\
\ \ \ \ \tilde \Delta_0=1-{a^2\ov U^2}\ ,
\eeq
\beq
d\Omega_3^2=d\psi_1^2+ \sin^2\psi_1 d\psi_2^2+\cos^2\psi_1 d\psi_3^2 \ .
\eeq

The theory describes fermions with anti-periodic boundary conditions
on the circle parameterized by $\tau $, which has radius $(2\pi T_H)^{-1}$
with
\beq
T_H={u_0\over\pi B} \ ,\ \ \ \ \ B\equiv {2u_0^3\over u_H(2 u_H^2-a^2)}\ ,
\label{tthh}
\eeq 
\beq
u_H^2=\frac{1}{2}( a^2+\sqrt{a^4+4u_0^4})\ .
\non
\eeq
For convenience, we have rescaled variables
by $U=(\rr )^{1/2}\ u\ ,\ a\to (\rr)^{1/2}\ a$.
At energies much lower than $T_H$ the theory should be effectively
$2+1$ dimensional (with $x_0=ix_3$ playing the role of time).
The gauge coupling of the $2+1$ dimensional field theory is given by
\beq
g_{\rm YM_3}^2=g^2_{\rm YM_4}T_H\ ,\ \ \ \ \  \  g^2_{\rm YM_4}=2\pi\ g_s\
{}.
\label{coup}
\eeq
\beq
\lambda\equiv {g_{\rm YM_3}^2N \over 2\pi }= g_s N \ {u_0\over \pi B}
\eeq
In this model Wilson loops exhibit an area-law behavior with string tension
\beq
\sigma ={1\over 2\pi }\sqrt{\rr }\ u_0^2=
\sqrt{\lambda B}\ u_0^{3/2}
\ .
\label{tenss}
\eeq
%The gauge coupling of the $2+1$ dimensional field theory is given by
%\beq
%g_{\rm YM_3}^2=g^2_{\rm YM_4}2\pi T_H\ ,\ \ \ \ \  \  g^2_{\rm YM_4}=2\pi 
%g_s\
%{}.
%\label{coup}
%\eeq
%\beq
%\lambda\equiv g_{\rm YM_3}^2N= (4\pi g_s N) {u_0\over B}
%\eeq
%In this model Wilson loops exhibit an area-law behavior with string tension
%\beq
%\sigma ={1\over 2\pi }\sqrt{\rr }\ u_0^2=
%{\sqrt{\lambda B}\over 2\pi  }\ u_0^{3/2}
%\ .
%\label{tenss}
%\eeq
This can be obtained by minimizing the Nambu-Goto action \cite{malda}, 
and it
is essentially given by the coefficient of $\sum_{i=1}^3 dx_i^2$ at the 
horizon
times $1/(2\pi )$ \cite{WittenAdsThermal,brand,rey}.
In the limit of large 't Hooft coupling, the light physical states
are the supergravity modes, whose
masses can be determined from the equations of motion of the
string theory effective action. In the next sections we calculate
the mass spectrum of the light physical states and of KK modes.

%%%%%%%%%%%%%%%%%%%%%%%%%%%%%%%%%%%%%%%%%%%
\subsection{Spectrum of glueball masses}
%%%%%%%%%%%%%%%%%%%%%%%%%%%%%%%%%%%%%%%%%

In order to find the spectrum of the $0^{++}$ glueball states
one has to consider the supergravity equation for the
dilaton mode $\Phi$ that couples to
the  operator $\tr F^2$
\beq
\del_\mu \sqrt{g}
g^{\mu\nu}\del_\nu \Phi=0
\comma
\eeq
evaluated in the above background.
For functions of the form $\Phi=\chi (U)\  e^{ik\cdot x} $, we obtain
\beq
\partial_u [ (u^4-u_0^4-a^2u^2)u\chi' (u)]+M^2u\chi (u)=0, \ \ \ M^2=-k^2\ .
\eeq
The eigenvalues of this equation can again be determined numerically. The
results are presented in Table \ref{tab:3ddil} and Figure \ref{fig:3ddil}.
Figure \ref{fig:3ddil} gives the dependence on $a$ of the mass ratio of the
first excited $0^{++}$ glueball state compared to the ground state
$0^{++}$. One obtains a very similar behavior to the case of
QCD$_4$, that is the mass ratio changes very little, and takes on its
asymptotic value quickly. The comparison to the available lattice
results \cite{Teper3d} are given in Table \ref{tab:3ddil}

\begin{table}[htbp]
\centering
\begin{tabular}{l|ccc}
state & lattice $N\to \infty$&
value for $a=0$ & value for $a\to \infty$ \\
 \hline
 $0^{++}$ & 4.065 $\pm$ 0.055 & 4.07 (input) & 4.07 (input) \\
 $0^{++*}$ & 6.18 $\pm$ 0.13  & 7.03 &  7.05 \\
 $0^{++**}$ & 7.99 $\pm$ 0.22 & 9.93 & 9.97 \\
 $0^{++***}$ & - &  12.82 & 12.87 \\
\end{tabular}
\parbox{5in}{\caption{Masses of the $0^{++}$ glueballs and their excited
states in QCD$_3$. The first column gives the lattice results
extrapolated to $N\to \infty$, the second column the supergravity
results for $a=0$ and the third column the supergravity limit $a\to \infty$.
The lattice results are in the units of the square root of the string tension.
The error given is statistical and does not include the systematic error.
\label{tab:3ddil}}}
\end{table}

\begin{figure}
\PSbox{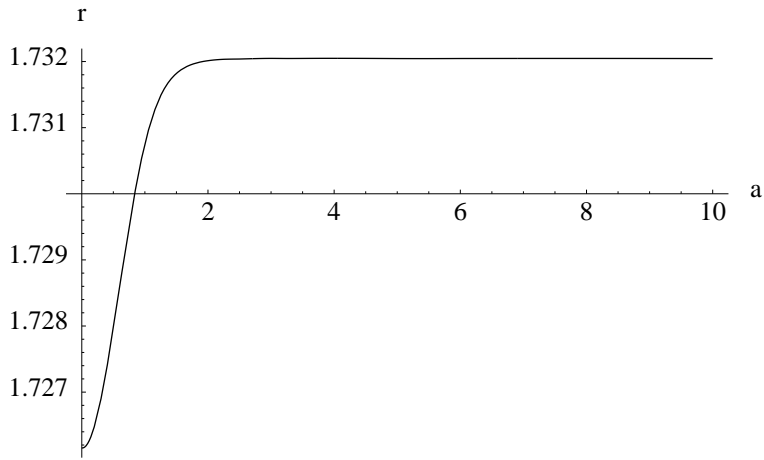 hscale=100 vscale=100 hoffset=50
voffset=-60}{13.7cm}{4.5cm}
\begin{center}
\parbox{5in}{\caption{The 
dependence on $a$ of the ratio $r=\frac{M_{0^{++*}}}{M_{0^{++}}}$
in QCD$_3$. One can see that the ratio is very stable
to changes in $a$, and reaches its asymptotic value quickly. $a$ is
given in units of $u_0$.
\label{fig:3ddil}}}
\end{center}
\end{figure}

%%%%%%%%%%%%%%%%%%%%%%%%%%%%%%%%%%%%%%%%%%%%

%%%%%%%%%%%%%%%%%%%%%%%%%%%%%%%%%%%%%%%%%%%
\subsection{Masses of KK states}
%%%%%%%%%%%%%%%%%%%%%%%%%%%%%%%%%%%%%%%%%

Just like in the case of QCD$_4$, we would like to analyze the
behavior of the masses of the different KK modes. We will find very similar
results: the KK modes wrapping the coordinate $\tau$ are decoupling
(even though a bit slower than in QCD$_4$), while the other KK modes
corresponding to states with angular momentum on the $S^5$ are not
decoupling in the supergravity limit.

First we consider the KK modes wrapping the compact $\tau$ direction.
Let us consider solutions to the Laplace equation
$\nabla^2\Phi=0$ of the form
\beq
\Phi=\chi(u)\ e^{ik\cdot x} e^{i\beta \tau}\ .
\eeq
The coordinate $\tau $ is periodic with period $T_H^{-1}$, where $T_H$ is given in 
eq.~(\ref{tthh}). Therefore
\beq
\beta = 2\pi  T_H\ n\ ,\
\eeq
where $n$ is an integer. Using the metric (\ref{3dmetric}) we find
\beq
 {\partial_u }((u^4-u_0^4-a^2u^2)u\chi'(u))=u
(-M^2+ n^24\pi^2 T_H^2 \frac{u^2-a^2}{u^2-a^2-\frac{u_0^4}{u^2}})\chi (u).
\eeq
One can see that just like in the case of QCD$_4$ there is an additional
positive contribution to the masses, which grows like $a^4$, therefore
the masses of these KK states should grow as $M^2_{KK}\approx a^4$.
Thus these
KK modes decouple from the spectrum, but slower than the corresponding
KK modes in QCD$_4$. The results of the numerical analysis are summarized
in Table \ref{tab:3dkk} and Figure \ref{fig:3dkk}.

\begin{table}[htbp]
\centering
\begin{tabular}{l|cc}
state &
value for $a=0$ & value for $a=4$ \\
 \hline
 $KK$ & 5.79  &  23.77 \\
 $KK^{*}$ & 8.64 &   24.63 \\
 $KK^{**}$ & 11.50 &  25.78 \\
 $KK^{***}$ & 14.36 & 27.19  \\
\end{tabular}
\parbox{5in}{\caption{Masses of the KK
modes wrapping the circle $\tau$ in
QCD$_3$, using the same normalization as in Table \protect\ref{tab:3ddil}.
The first column gives the masses
for $a=0$ while the second the
masses for $a=4$.
Note that these states decouple quickly
from the spectrum  even in the supergravity approximation.\label{tab:3dkk}}}
\end{table}

\begin{figure}
\PSbox{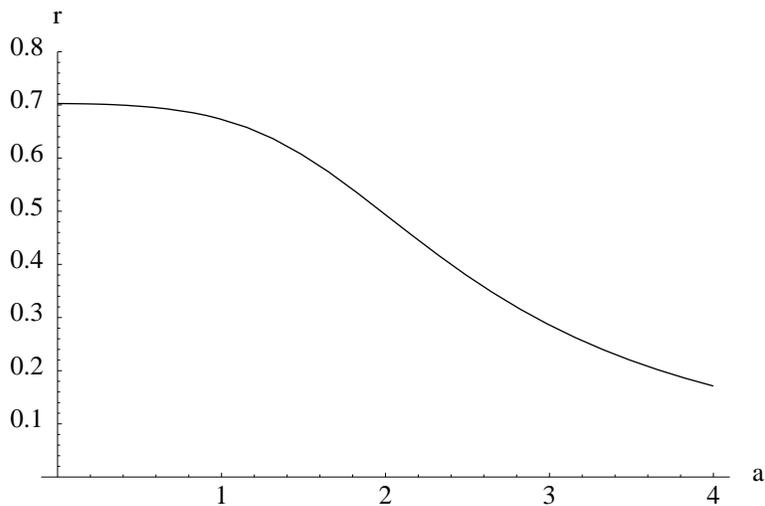 hscale=100 vscale=100 hoffset=50
voffset=-60}{13.7cm}{4.5cm}
\begin{center}
\parbox{5in}{\caption{The dependence on $a$ 
of the ratio $r=\frac{M_{0^{++}}}{M_{KK}}$
of the lowest $0^{++}$ glueball
state compared to the KK
states wrapping the $\tau$ circle in units where $u_0=1$.
These KK modes decouple from the spectrum in
the supergravity approximation very quickly.
\label{fig:3dkk}}}
\end{center}
\end{figure}

Next we analyze the KK modes which correspond to states with 
angular momentum $l=1$ on $S^5$ in the $a=0$ case. For $a=0$ these
states have been examined in \cite{ORT}, and found to be non-decoupling
in the supergravity limit and including the lowest order $\alpha'$ corrections.
Here we repeat this analysis and find (just like in the case of QCD$_4$)
them to be non-decoupling even in the $a\to \infty$ case, in the supergravity
limit.

In order to do the analysis of these KK modes one needs to find the
explicit form of the spherical harmonics.
The spherical harmonics on $S^d$ can be constructed in the following way.
One takes $S^d$ embedded in $R^{d+1}$, and expresses the
Cartesian coordinates $y_i$
in terms of the angles.
Then the spherical
harmonics are just the functions $C^{i_1,...,i_k} y_{i_1}....y_{i_k}$,
where $C$ is a symmetric traceless tensor \cite{Seiberg}.
This way, the simplest non-trivial spherical
harmonic is just the coordinate $y_i$ itself.
In the case of our QCD$_3$ theory, we actually have to use the
``spheroidal coordinates" $y_i$ given in \cite{russo} on page 9.
Thus one looks for solutions of the dilaton equation of the form
$f(u) e^{ik\cdot x} y_i,\ \  i=1,2,3,4,5,6$.
For $a=0$ the isometry group of $S^5$ is $SO(6)$, and the
$l=1$ KK mode is in the representation ${\bf 6}$ of $SO(6)$.
Introducing the angular momentum
$a$ breaks $SO(6)$ to $SO(4)\times U(1)\times U(1)$, and the
${\bf 6}$ decomposes into ${\bf 4}+{\bf 1}+{\bf 1}$.
These states satisfy different eigenvalue equations.
For $i=1,2$ (the two singlets are degenerate) the equation one gets
is
\beq
\partial_u [u(u^4-a^2u^2-1)f'(u)]-f(u)\left( k^2 u +
\frac{5u^3+4a^2u^5-5u^7}{1+a^2u^2-u^4}\right) =0.
\eeq
Note that for $a=0$ this indeed reduces to the equation given in
\cite{ORT} for $l=1$. We have numerically solved this equation, and find
that the mass of these KK modes is growing slightly, until it 
becomes 
degenerate with the first excited state of the $0^{++}$ glueball. Thus
it does not decouple from the spectrum in the supergravity limit.
The results are summarized
Table \ref{tab:3dnon1} and Figure \ref{fig:3dnon1}.

\begin{table}[htbp]
\centering
\begin{tabular}{l|cc}
state &
value for $a=0$ & value for $a\to \infty$ \\
 \hline
 $KK$ & 5.27 &  7.05 \\
 $KK^{*}$ & 8.29 &   9.97 \\
 $KK^{**}$ & 11.23 &  12.87 \\
 $KK^{***}$ & 14.14 & 15.76  \\
\end{tabular}
\parbox{5in}{\caption{Masses of the KK
modes corresponding to the two degenerate singlet
pieces of the $l=1$ sextet of the
original $SO(6)$ isometry in
QCD$_3$, using the same normalization as in Table \protect\ref{tab:3ddil}.
The first column gives the masses
for $a=0$ while the second the
masses in the $a\to \infty$ limit. Note that these states do not decouple
from the spectrum  in the supergravity approximation.\label{tab:3dnon1}}}
\end{table}

\begin{figure}
\PSbox{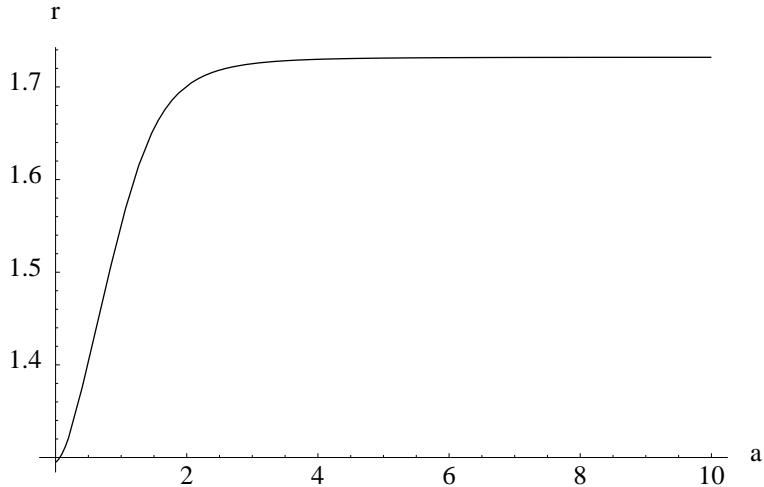 hscale=100 vscale=100 hoffset=50
voffset=-60}{13.7cm}{4.5cm}
\begin{center}
\parbox{5in}{\caption{The dependence of the ratio $r=\frac{M_{KK}}{M_{0^{++}}}$
of the KK
modes corresponding to the two singlet $l=1$ states
compared to the lowest $0^{++}$ glueball
state on the parameter $a$ in units where $u_0=1$.
This KK mode does not decouple from the spectrum in
the supergravity approximation even in the $a\to \infty$ limit, even
though it increases slightly and becomes exactly degenerate with the first
excited glueball state $0^{++*}$.
\label{fig:3dnon1}}}
\end{center}
\end{figure}

For the other 4 KK states which are in the ${\bf 4}$ of $SO(4)$
one finds the equation
\begin{equation}
\partial_u [u(u^4-a^2u^2-1)f'(u)] -f(u)(k^2 u+5u^3-3a^2u)=0\ ,
\eeq
which for $a=0$  reproduces the equation in \cite{ORT}
with $l=1$.
The additional mass term is now negative, which actually makes
these KK modes lighter in the $a\to \infty$ limit than
for $a=0$. However, they are still of the same order (and slightly
heavier) than the $0^{++}$ glueballs. The numerical results
for this state are summarized in Table \ref{tab:3dnon2} and
Figure \ref{fig:3dnon2}.

\begin{table}[htbp]
\centering
\begin{tabular}{l|cc}
state &
value for $a=0$ & value for $a\to \infty$ \\
 \hline
 $KK$ & 5.27 &  4.98 \\
 $KK^{*}$ & 8.29 &   8.14 \\
 $KK^{**}$ & 11.23&  11.15 \\
 $KK^{***}$ & 14.14& 14.10   \\
\end{tabular}
\parbox{5in}{\caption{Masses of the KK
modes corresponding to the quartet piece of the $l=1$ sextet of the
original $SO(6)$ in
QCD$_3$, using the same normalization as in Table \protect\ref{tab:3ddil}.
The first column gives the masses
for $a=0$ while the second the
masses in the $a\to \infty$ limit. Note that these states actually get
lighter
from $a=0$ to $a=\infty$ in the supergravity approximation.\label{tab:3dnon2}}}
\end{table}

\begin{figure}
\PSbox{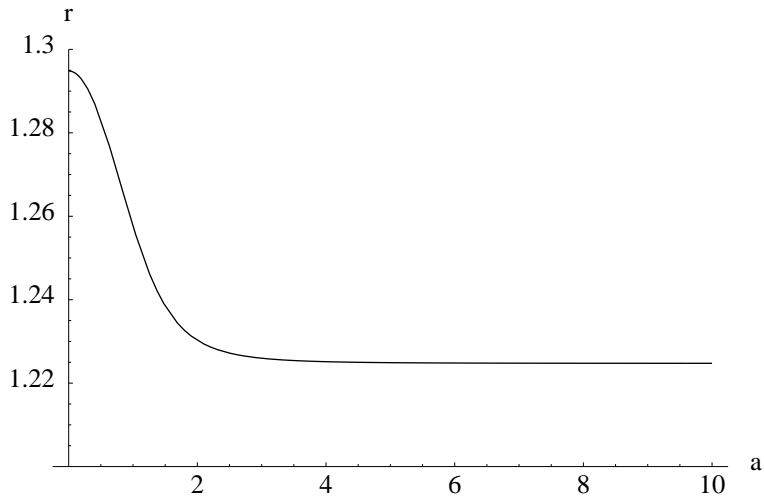 hscale=100 vscale=100 hoffset=50
voffset=-60}{13.7cm}{5.5cm}
\begin{center}
\parbox{5in}{\caption{The dependence of the ratio $r=\frac{M_{KK}}{M_{0^{++}}}$
of the KK
states compared to the lowest $0^{++}$ glueball
state on the parameter $a$ in units where $u_0=1$.
This KK mode does not decouple from the spectrum in
the supergravity approximation even in the $a\to \infty$ limit,
they instead get even slightly lighter than for $a=0$.
\label{fig:3dnon2}}}
\end{center}
\end{figure}

%%%%%%%%%%%%%%%%%%%%%%%%%%%%%%%%%%%%%%%%%%%
\subsection{Free energy and Gluon condensation}
%%%%%%%%%%%%%%%%%%%%%%%%%%%%%%%%%%%%%%%%%

{}From the rotating D3 brane metric (see Eq.~(3.16) in \cite{russo}), one can
find the following
formulas for the thermodynamic variables:
\beqar
M_{\rm ADM}&=&{V_3V(\Omega_5)\ov 4\pi G_N} {5\over 4}
2m (1+{4\ov 5}\sinh^2\alpha )\ ,\  \ \ \ V(\Omega_5)=\pi ^3\ ,
\label{mmt}
\\
S &=& {V_3V(\Omega_5)\ov 4 G_N} 2mr_H \cosh\alpha \ ,
\label{mmv}\\
J_H &=&{V_3 V(\Omega_5)\over 4\pi G_N}
ml\cosh\alpha \ ,
\label{mmx}
\eeqar
\beq
T_H={r_H(2r_H^2+l^2)\over 4\pi m \cosh\alpha }\ ,\ \ \ \ \ \
\Omega_H={l r_H^2\over 2m\cosh\alpha }
\ ,\
\label{oott}
\eeq
\beq
G_N={\kappa^2_{10}\ov 8\pi }= 8 g_s^2\pi^6 (\alpha')^4\ ,
\label{ggnn}
\eeq
where
\beq
2 m \cosh\alpha\sin\alpha =4\pi g_sN {\alpha'}^2\ .\
\eeq
One can check that they satisfy the first law of black hole thermodynamics
Eq.~(\ref{fll}).
In the limit $\alpha '\to 0$ (rescaling variables as in (\ref{resc})~),
we get
\beq
E\equiv M_{\rm ADM}-M_{\rm ext}=
{3\over 8\pi^2} V_3\ N^2\ u_0^4 \ ,
\label{wuno}
\eeq
\beq
T_H={u_0\over\pi B} \
\ ,\ \ \ \ \ S={1\over 2\pi} V_3\ N^2\ u_H u_0^2 \ ,
\label{wdos}
\eeq
\beq
\Omega_H=i {a u_H^2\over u_0^2}
\ ,\ \ \ \ \  J_H=i {1\over 4\pi^2}V_3\ N^2\ a u_0^2 \ ,
\label{wtres}
\eeq
\beq
u_H^2-a^2={u_0^4\over u_H^4}\ ,\ \ \ \ \ B= {2u_0^3\over u_H(2u_H^2-a^2)}\ .
\eeq
The free energy is then given by
\beq
F=E-T_HS- \Omega_H J_H=
-V_3 {N^2u_0^4\over 8\pi ^2}\  .
\label{wfree}
\eeq
This gives for the gluon condensate the following expression:
\beq
\langle {1\over 4g_{\rm YM}^2 }\tr\  F^2_{\mu\nu}(0) \rangle =
-{F\over V_3T_H}={1 \over 8\pi} N^2 B u_0^3\ .
\label{ggl}
\eeq
In terms of the Yang-Mills string tension, this is
\beq
\langle {1\over 4g_{\rm YM}^2 }\tr\  F^2_{\mu\nu}(0) \rangle =
{1\over 8\pi } {N^2\over\lambda}  \ \sigma^2\ .
\label{ggl2}
\eeq
%In terms of the Yang-Mills string tension, this is
%\beq
%\langle {1\over 4g_{\rm YM}^2 }\tr\  F^2_{\mu\nu}(0) \rangle =
%{\pi\over 2} {N^2\over\lambda}  \ \sigma^2\ .
%\label{ggl2}
%\eeq
We  find again that supergravity predicts that
the gluon condensate is proportional to
$N^2/\lambda $ times the string tension squared.
The result expressed in terms of the string tension is thus
independent of $a$.

%%%%%%%%%%%%%%%%%%%%%%%%%%%%%%%%%%%%%%%%%%%
\section{Conclusions}
%%%%%%%%%%%%%%%%%%%%%%%%%%%%%%%%%%%%%%%

In this paper we investigated quantitative aspects
of  large $N$ $SU(N)$ Yang-Mills theory in
three and four dimensions using a one-parameter
family of
supergravity models related to  non-extremal rotating
D-branes.
The  new feature of this class of models is
the decoupling of the KK
modes associated with the compact D-brane coordinate
as the angular momentum
parameter is increased.
The mass ratios for ordinary glueballs were found to be very stable
against this variation. While the mass ratios 
 of the $0^{++}$ glueballs change 
only slightly compared to the case with zero angular momentum, there is a 
substantial change in the mass ratios 
 of  $0^{-+}$, $0^{-+*}$  
given in Eqs. (\ref{que}), (\ref{que2}), which for large $a$ are in better 
agreement with the lattice values than for $a=0$.

It is worth emphasizing that the ratio $a/u_0$ should be large
enough to have $M_{KK}\gg M_{\rm glueball}$, but not
infinite, since  there are 
also  string states winding around the compact D-brane coordinate with 
masses of order $\sigma R_0$ that should decouple, i.e. $M_{\rm wind}\gg M_{\rm
glueball}$.
This requires $\lambda u^8_0/a^8 \gg 1$, which is consistent with the
condition that curvature invariants are small compared to the string scale
 \cite{russo}.
In general, for any given ratio $a/u_0$ which is large enough to decouple
KK states from the low-energy physics,
it is possible to  choose $\lambda $ sufficiently large
so that string winding states also decouple.

We have  found that the ($SO(3)$ or $SO(4)$) non-singlet KK modes
with vanishing $U(1)$ charge in the compact D-brane coordinate
do not decouple in this class of models.
One can hope that those KK  modes may decouple in a model with
 more angular momenta (since there is room to take other limits).
In this case the isometry group of the internal space is smaller. For example,
in ${\rm QCD}_3$, for $a=0$ it is given by $SO(6)\times U(1)$, whereas
for $a\neq 0$ it is $SO(4)\times
U(1)\times U(1)$. The isometry group of the model with the maximum number of
angular momenta
only contains $U(1)$ factors.  This is consistent
with the fact that in pure QCD there can  only be  singlets of the
original R-symmetry.

We have also found some features  which seem to be universal, i.e.
which do not depend on the extra supergravity parameter.
In particular, both in ${\rm QCD}_3$ and ${\rm QCD}_4$
supergravity gives a gluon condensate of the form ${N^2\ov \lambda} \sigma^2$,
with a coefficient which is the same for
all models parametrized by $a$.
Another  feature that seems to be common to all supergravity models
 is a topological susceptibility of the form
 $\lambda \sigma^2 $,
 with a coefficient which  is independent of $N$ but depends on $a/u_0$.
This result suggests that in the  regime $\lambda\gg 1 $ the $\eta' $
particle of ${\rm QCD}_4$ with $N=3$ is much heavier than other mesons
(whose masses are proportional to the string tension).

%%%%%%%%%%%%%%%%%%%%%%%%%%%%%%%%
\section*{Acknowledgements}
%%%%%%%%%%%%%%%%%%%%%%%%%%%%%%%%

We thank M. Dine, A. Hashimoto, K. Hori, K. Jansen, C. Morningstar,
H. Ooguri, M. Peardon, and A. Zaffaroni
for useful discussions.
C.C. is a research fellow of the Miller
Institute for Basic Research in Science.
C.C. and J.T. are supported in part
the U.S. Department of Energy under Contract DE-AC03-76SF00098, and in part
by the National Science Foundation under grant PHY-95-14797.
The work of J.R. is supported by the European
Commission TMR programme grant  ERBFMBI-CT96-0982.
%%%%%%%%%%%%%%%%%%%%%%%%%%%%%%%%%%%%%%%%%%%%% 
%%%%%%%%%%%%%%%%%%%%%%%%

\end{document}